\documentclass[twocolumn,aps,prl,showpacs,raggedbottom,nobalancelastpage,amssymb,superscriptaddress]{revtex4-1}

\usepackage{graphicx}
\usepackage{amsmath}
\usepackage{amssymb}
\usepackage{bm}
\usepackage[usenames]{color}
\usepackage{amsfonts}
\usepackage{appendix}
\usepackage{dsfont}
\usepackage[breaklinks=true,colorlinks=true,linkcolor=blue,urlcolor=blue,citecolor=blue]{hyperref}
\graphicspath{{Figures/}}

\def\be{\begin{equation}}
\def\ee{\end{equation}}

\begin{document}
\title{Superfluidity of Light and its Break-Down in Optical Mesh Lattices}
\author{Martin Wimmer}
\author{Monika Monika}
\address{Institute of Condensed Matter Theory and Optics Friedrich-Schiller-University Jena, Max-Wien-Platz 1,
D-07743 Jena, Germany}

\author{Iacopo Carusotto}
\address{INO-CNR BEC Center and Dipartimento di Fisica, Universit\`{a} di Trento, I-38123 Povo, Italy}

\author{Ulf Peschel}
\address{Institute of Condensed Matter Theory and Optics Friedrich-Schiller-University Jena, Max-Wien-Platz 1,
D-07743 Jena, Germany}

\author{Hannah M. Price}
\address{School of Physics and Astronomy, University of Birmingham,
Edgbaston Park Road, B15 2TT, West Midlands, United Kingdom}

\begin{abstract}
Hydrodynamic phenomena can be observed with light thanks to the analogy between quantum gases and nonlinear optics. In this Letter, we report an experimental study of the superfluid-like properties of light in a (1+1)-dimensional nonlinear optical mesh lattice, where the arrival time of optical pulses plays the role of a synthetic spatial dimension. A spatially  narrow defect at rest is used to excite sound waves in the fluid of light and measure the sound speed. The critical velocity for superfluidity is probed by looking at the threshold in the deposited energy by a moving defect, above which the apparent superfluid behaviour breaks down.  Our observations establish optical mesh lattices as a promising platform to study fluids of light in novel regimes of interdisciplinary interest, including non-Hermitian and/or topological physics.
\end{abstract}
\maketitle

The last decades have witnessed an impressive development of new conceptual links between the apparently disconnected fields of nonlinear optics and many-body physics of quantum gases~\cite{RevModPhys.85.299}. The formal analogy between the paraxial light propagation in nonlinear media and the Gross-Pitaevskii equation of dilute Bose-Einstein condensates was first noticed in the 1970s and immediately suggested the transfer of concepts such as superfluidity and quantized vortices to optical systems~\cite{Pomeau1993,Arecchi1991,Swartzlander:PRL1992,Frisch:PRL1992,Staliunas_book}. This connection was further revived with the experimental observation of Bose-Einstein condensation in exciton-polariton gases in semiconductor microcavities~\cite{kasprzak2006bose}, which triggered a strong interest from both theory and experiment to address basic features of condensates such as superfluidity, hydrodynamics and topological excitations~\cite{Amo:NPhys2009,Amo:2011Science,Nardin:2011NatPhys} in the new context of fluids of light. 

\begin{figure}[!]
	\includegraphics[width=1\linewidth]{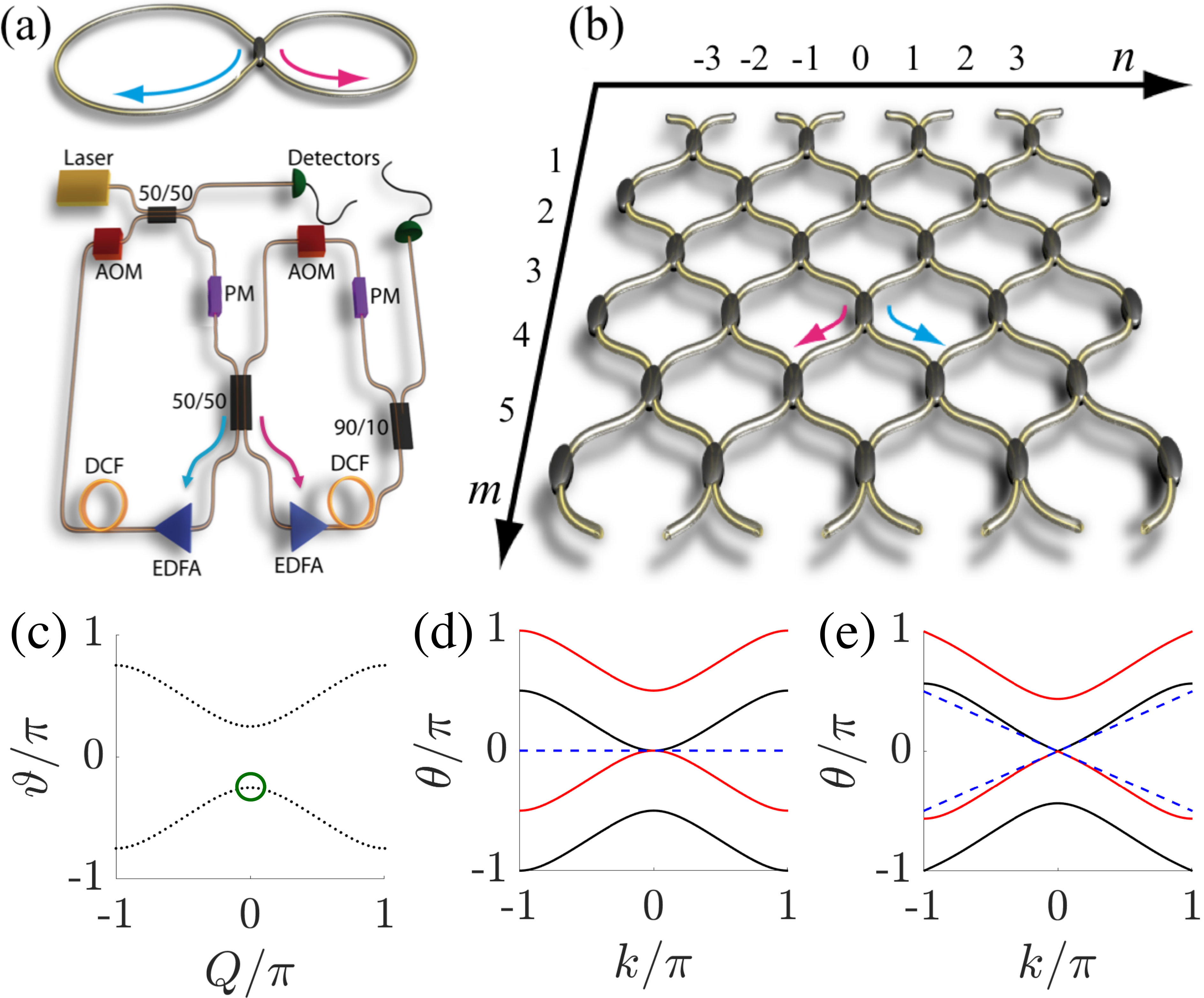}
	\caption{(a) Optical pulses propagating in two nonlinear, coupled fiber loops of slightly different lengths, are used to explore nonlinear light evolution in the $(1+1)$D lattice, shown schematically in (b). In this mapping, the light intensity is a function of the discrete position in the lattice, $n$, and evolves with respect to the discrete time-step, $m$. Completing a round-trip in the short (long) loop in the real system in (a) corresponds to travelling from northeast (northwest) to southwest (southeast) in the effective lattice in (b). Acousto-optical modulators (AOM) and erbium doped fiber amplifiers (EDFA) are used to compensate for losses. A phase modulator (PM) in each loop allows us to induce arbitrarily designed space- and time-dependent potentials. (c) The corresponding photonic bands in the linear ($\Gamma=0$) regime. (d)\&(e) The Bogoliubov dispersions \eqref{eq:bog} on top of a condensate located  at $Q=0$ in the lower band [circle in (c)]  for (d) linear and  (e) nonlinear ($\Gamma I_0=0.2$) systems. The slope of the straight blue dashed line  indicates the speed of sound \eqref{eq:sound}. The red/black color of each curve indicates the positive/negative value of the band's Bogoliubov norm.
	 }\label{setup}	
\end{figure}

In this work, we introduce a new platform for studying fluids of light by using classical light in a so-called optical mesh lattice. The idea is to encode one (or even more~\cite{schreiber2,muniz}) discrete synthetic spatial dimensions in the arrival time of optical pulses that propagate along coupled optical fiber loops~\cite{schreiber2010photons,regensburger2012parity,miri2012optical}. This allows for the application of arbitrary dynamical potentials to the fluid of light and for the measurement of its evolution in real time with key advantages over traditional systems. In contrast to semiconductor microcavities~\cite{RevModPhys.85.299}, our system provides great flexibility in the design of different lattice geometries with no fabrication effort and naturally offers site-resolved access to the temporal dynamics of the fluid without the need for sophisticated ultrafast optics tools. In contrast to bulk nonlinear systems~\cite{vocke2015experimental,fontaine2018observation,michel2018superfluid}, the nonlinearity stems from the power dependent propagation constant of optical fibers and is thus fully controllable with standard opto-electronic tools. Although the potential of such optical mesh lattices for studies of linear and nonlinear optics has been demonstrated in many works on, e.g., $PT$-symmetric physics~\cite{regensburger2012parity,miri2012optical, wimmer2013optical, wimmer2015observation:scirep, wimmer2015observation, muniz} and topological effects~\cite{Chen2018, Chalabi2019,bisianov,weidemann2020topological}, here we report their first use as a platform to investigate fluids of light. In the future, it will be of great interest to further exploit the flexibility of optical mesh lattices to study fluids of light in topological and/or non-Hermitian regimes that are not straightforwardly realized in standard platforms, e.g. combining interactions with synthetic magnetic fields, topological lattices~\cite{RevModPhys.91.015006}, and complex gain/loss distributions~\cite{ashida2020non,lledo2021spontaneous}.

{\it The experimental set-up and the theoretical model --} In our experiments, an optical mesh lattice is realised using a time-multiplexing scheme based on two optical fiber loops of average length $\bar{L}\!=\!4$~km, which have a small length difference $\delta L \!=\! 50$~m and which are coupled by a $50/50$ beamsplitter, as shown in Fig.~\ref{setup}(a). A light pulse injected into one loop is split by the coupler into two pulses: one circulating in the longer and one in the shorter loop. After a round trip, the two pulses arrive back at the coupler but now with a relative time-delay due to the small length difference between the loops. Each pulse is again split into two by the beamsplitter, which, after many round-trips, eventually leads to the generation of a pulse train over time, as reviewed in the Supplemental Material~\cite{supmat}. 

There is a clear separation of time-scales between the average round-trip time $\bar{T}$ and the relative time-delay $\Delta T$ as $\bar{T}\gg \Delta T$. Thus, for not too many round trips $m$ ($m\, \Delta T < \bar{T}$) the arrival time of each pulse at the beamsplitter can be unambiguously expanded as $t\!=\! m  \bar{T} + n \Delta T/2 $, where the integer $m$ denotes the total number of round trips for each pulse and the integer $n$ counts how many more round trips were made in the long rather than in the short loop~\cite{supmat}. As the light propagates, the assigned integer $m$ keeps increasing to count each successive round trip, while the integer $n$ increases or decreases by one after each round trip depending on which loop is traversed. The integer $m$ is interpreted as a discrete time-step, and $n$ as a discrete position index along a ``synthetic spatial dimension", leading to the effective $(1+1)$D lattice shown in Fig.~\ref{setup}(b)~\cite{schreiber2010photons,regensburger2012parity,miri2012optical}. 
Even though this dynamics is intrinsically discrete in both space and time, for spatiotemporally slow fields it has an intriguing continuum limit of a Dirac model~\cite{supmat}.

To explore nonlinear phenomena, dispersion compensating fiber (DCF) spools form part of the long and the short loops. These DCFs have a narrow core size and thus a high effective nonlinear coefficient~\cite{gruner2005dispersion} of approximately 7/km/W, leading to significant nonlinear effects in our experiment~\cite{agrawal2013nonlinear}. We use DCF spools with a length of approximately 4~km and employ peak powers on the order of 100~mW. To sustain such high power, erbium-doped fiber amplifiers are used to compensate for round-trip losses. 
As reviewed in the Supplementary Material~\cite{supmat}, the significant Kerr nonlinearities lead to a power-dependent phase-shift for the propagating light pulses, as captured by~\cite{regensburger2011photon,Navarrete,wimmer2013optical} 
\begin{eqnarray}
u_n^{m+1} &=& \frac{1}{\sqrt{2}} \left( u_{n+1}^m e^{i \Gamma |u_{n+1}^m|^2} + i v_{n+1}^m e^{i \Gamma |v_{n+1}^m|^2} \right) e^{i \varphi_n^m} , \nonumber \\
v_n^{m+1} &=& \frac{1}{\sqrt{2}} \left( v_{n-1}^m e^{i \Gamma |v_{v-1}^m|^2} + i u_{n-1}^m e^{i \Gamma |u_{n-1}^m|^2} \right) e^{i \phi_n^m}, \label{eq:nonlinear}
\end{eqnarray}
where $\Gamma>0$ quantifies the effect of the nonlinear refractive index per round trip, corresponding to a negative interaction energy in the quantum fluids language. Here, $u_n^m$ and $v_n^m$ denote the amplitudes of the pulses incident on the beamsplitter from the short and long loops respectively. The time-dependent linear phase-shifts $\varphi_n^m$ and $\phi_n^m$ are externally controlled through phase modulators inserted in each loop, and can be used, for example, to imprint defects or trapping potentials on the fluid of light. 

As the equations are periodic 
in both position $n$ and time-step $m$, they can be solved in the linear regime ($\Gamma\!=\!0$) using a Floquet-Bloch ansatz~\cite{miri2012optical,forthcoming}. This gives two bands, which are $2\pi$ periodic in both the propagation constant or ``quasi-energy" $\vartheta$ and the ``Bloch momentum'' $Q$ along the 1D lattice, with a dispersion relation $\cos \vartheta=\cos Q/ \sqrt{2}$ as shown in Fig.~\ref{setup}.
	
{\it Sound excitations -- } The linear dynamics of small excitations, 
$\delta u_n^m$ and $\delta v_n^m$,
on top of a strong field, $u_n^m$ and $v_n^m$, can be described within the Bogoliubov theory of dilute Bose-Einstein condensates~\cite{pitaevskii2016bose,CastinLectures}. For simplicity, we consider the case where
the fluid of light is spatially uniform with the same amplitude $\sqrt{I_0}$ in both loops and is initially at rest in the $m,n$ coordinate system.
This
corresponds to a condensate of density $I_0$ in the lower band-eigenstate at $Q=0$ marked by a circle in Fig.~\ref{setup}(c).  Analytical manipulations lead to the Bogoliubov dispersion relation~\cite{martinthesis, forthcoming}
\begin{eqnarray}
\cos \theta &&= \frac{1}{2} \Gamma I_0 \cos k + \frac{1}{2} \cos k \nonumber \\
&&\pm \frac{1}{2} \sqrt{(\Gamma^2 I_0^2 + 2 \Gamma I_0 + 1) \cos^2 k + 2 \sin^2 k - 4 \Gamma I_0} , \qquad \label{eq:bog}
\end{eqnarray}
where $\theta$ and $k$ are respectively the quasi-energy and the Bloch momentum (both of period $2\pi$) of the small perturbation. 
This dispersion has two positive and two negative branches as shown in Fig.~\ref{setup}(d) for the linear ($\Gamma I_0 =0$) and (e) a weakly nonlinear ($\Gamma I_0 =0.2$)
case.

At small momenta, this Bogoliubov dispersion 
 can be expanded as $\theta (k) = \tilde{v}_{{S}} |k| +  \mathcal{O} (|k|^3)$, indicating that the long wavelength excitations are phonon-like, with a speed of sound, 
\begin{eqnarray}
 \tilde{v}_{{S}} =\sqrt{\Gamma I_0/ (1- \Gamma I_0)}, \label{eq:sound}
 \end{eqnarray}
 %
that grows for increasing nonlinearity faster than the usual square-root dependence of a simple Gross-Pitaevskii superfluid~\cite{pitaevskii2016bose}. Note that the positive sign of the nonlinear refractive index in the considered DCFs forced us to work with a negative-mass photonic band [i.e. the top of the lower band in Fig.~\ref{setup}(c)] to avoid dynamical modulational instabilities. 
 An upper bound to the sound speed is imposed by further dynamical instabilities of different nature that are found for $\Gamma I_0 \!> \!0.5$ 
 ~\cite{forthcoming}.

\begin{figure}[!]
	\includegraphics[width=\linewidth]{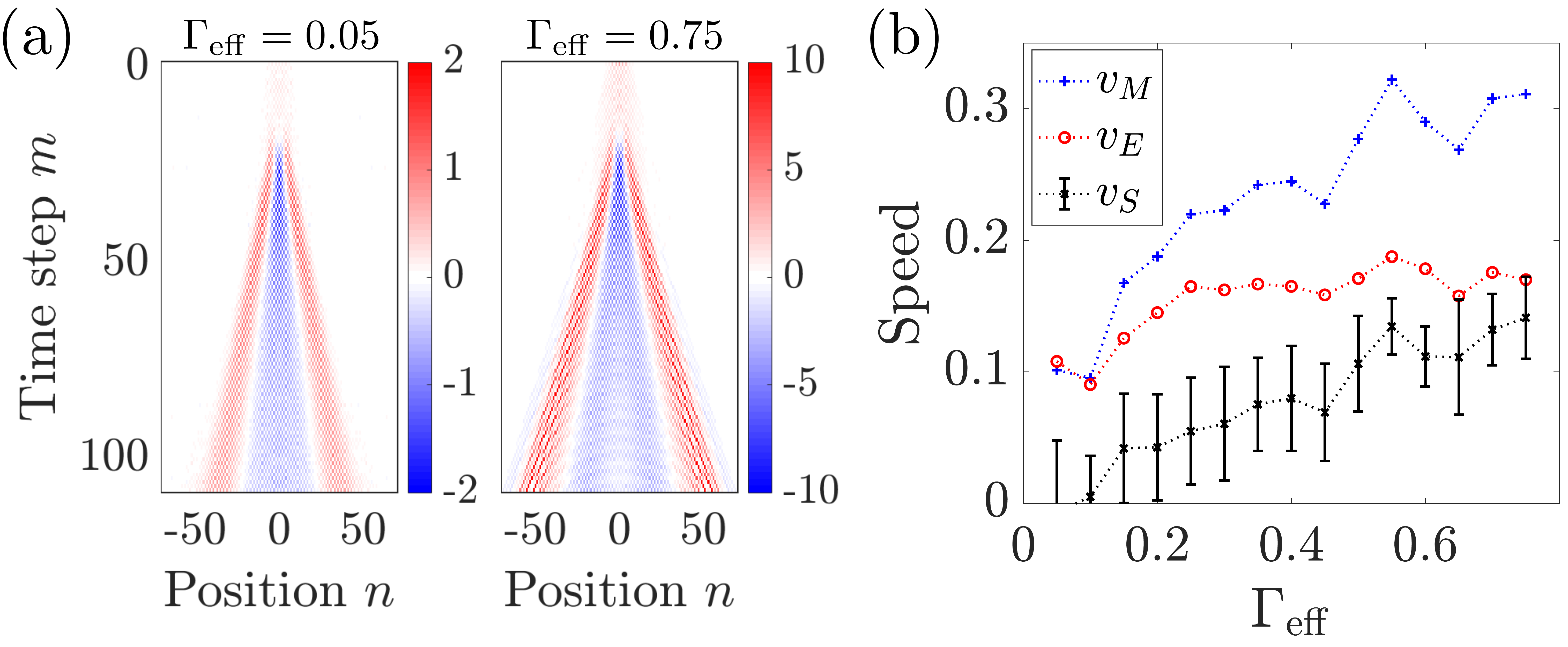}
	\caption{(a) Experimentally observed propagation of 
	perturbations induced by imposing  a short and localized phase defect onto
	a Gaussian fluid of light for two different values of the effective initial nonlinearity, $\Gamma_{\rm{eff}}\propto\Gamma I_0$. The colorscale represents the differential intensity, $\Delta I$, between the perturbed and unperturbed experiments. All parameters are  stated in the main text. 
	 (b) The estimated speed of sound $v_S=v_M-v_E$ grows from around zero with increasing nonlinearity. Also plotted are $v_M$, the average speed of the innermost minima at late times obtained by fitting the results in (a), and $v_E$, the average expansion speed of the unperturbed fluid of light at the minima positions, from which the sound speed was extracted. For clarity, errorbars for $v_M$ and $v_E$ are shown in the Supplemental Material~\cite{supmat}. 
	 Lines are included as a guide to the eye.
	 }\label{soundspeed}	
	\end{figure}

{\it Probing the speed of sound --} 
As a first step, we investigate the propagation of spatially narrow excitation pulses on top of a wide field. We initially prepare the optical field at rest with a wide Gaussian profile  corresponding to an equal amplitude in each loop at $m\!=\!0$: $u_n^0 \!= \!v_n^0\!=\! \sqrt{I_0} 
e^{- n ^2 / \sigma_g^2}$, where $\sigma_g\simeq 8$. Changing the input pulse power between experimental runs allows us to study the effect of nonlinearity in detail.
Note that according to the notation introduced in \eqref{eq:nonlinear} only every second lattice site is occupied as the effective lattice has diamond-like connectivity [Fig~\ref{setup}(b)]. 

To perturb this wide fluid of light, we rapidly turn on and off a strong and spatially localized defect potential
$\varphi_d$, imprinted via the time-dependent phase shifts in \eqref{eq:nonlinear} as
\begin{eqnarray}
  \varphi_{n}^m = \phi_{n}^m  = \Phi_d\equiv \varphi_d e^{-(n - n_d)^2 /  \sigma_n^2} e^{-(m - m_d)^2 / \sigma_m^2} \label{eq:defectform}
\end{eqnarray}
where $\varphi_d$ is the defect amplitude, $\sigma_n$ ($\sigma_m$) is related to the defect width with respect to position (time) and $n_d$ ($m_d$) marks the position (time) of the defect.
We choose $\varphi_d = \pi /10$ to ensure a sufficient contrast between the perturbed and unperturbed signal, and $ \sigma_n=1$ and $\sigma_m=2$
so that the defect is short enough to excite low-momentum perturbations. The defect peak amplitude is at $m_d=20$, and the propagation is observed over a long time until $m_{\rm{max}}=110$. 

The propagation dynamics of the emitted waves is measured by the differential intensity $\Delta I= I_{{\text{pert.}}} -I_{{\text{unpert.}}}$, where $I_{{\text{pert.}}}=|u_n^m|^2+|v_n^m|^2$ in the perturbed experiment, and similarly for $I_{{\text{unpert.}}}$ in the unperturbed case. This is plotted in Fig.~\ref{soundspeed}(a) for two different values of the effective nonlinearity: $\Gamma_{\text{eff}} \propto \Gamma I_0$, where $I_0$ is the initial light intensity which we can control. As can be seen, the defect emits a train of excitations which spread out over time~\footnote{The small differential intensity that is visible also before the defect is applied is due  to small deviations in the preparation of the initial state.}. We expect the slowest strong emission to be that associated with sound waves, and so we fit $\Delta I(n)$ at each time-step to extract the position of the minima~\cite{supmat}. The corresponding average speed, $v_M$, of the minima at late times is plotted in Fig.~\ref{soundspeed}(b). 

In order to extract the value of the speed of sound, we need to take into account the fact that the underlying fluid of light is also itself expanding during the experiment, dragging all excitations along with it. Hence, we numerically estimate a local expansion speed by applying a discretized continuity equation to the unperturbed evolution~\cite{supmat}. The corresponding average expansion speed, $v_E$, at the position of the minima at late times is plotted in Fig.~\ref{soundspeed}(b). By comparing these speeds, we finally obtain the estimate for the speed of sound $v_S = v_M-v_E$ plotted in Fig.~\ref{soundspeed}(b). Due to experimental uncertainties, we did not fit our data to the analytical sound speed (\ref{eq:sound}), but our results clearly show the expected qualitative trend as we measure $v_S$ as being close to zero at low power, and then rising with increasing nonlinearity. One may argue that the local light intensity and hence the local nonlinearity and speed of sound can vary over time. However, as we show in the Supplemental Material~\cite{supmat}, the local light intensity remains relatively stable because the amplifiers largely compensate for any intensity drops. The alternative way of plotting the data as a function of the averaged unperturbed local intensity at the position of the minimum exhibits the same qualitative features as Fig.~\ref{soundspeed}(b)~\cite{supmat}.

{\it Friction on moving defect --} A key advantage of the nonlinear optical mesh lattice is that we can
probe in the same set-up the response to defects which are moving 
at arbitrary speeds along arbitrary trajectories. One definition of superfluidity~\cite{Leggett} is the existence of a non-zero critical velocity below which a weak, uniformly moving defect will not excite permanent waves in the fluid. This critical velocity can be predicted by means of  the so-called Landau criterion, and for a Gross-Pitaevskii superfluid corresponds to the speed of sound~\cite{pitaevskii2016bose}. 

The situation is slightly more subtle in a lattice geometry, for which the Landau criterion predicts a vanishing critical velocity, as the $k$-periodicity of the Bogoliubov dispersion allows any moving defect, to emit excitations at larger momentum. In this sense, a lattice system can never truly be a superfluid, but in practice the excitation efficiency for such ``Umklapp waves" is negligibly small~\cite{forthcoming}. On this basis, we restrict our attention to low-transferred-momentum processes with $-\pi \leq \Delta k < \pi$, for which the Landau criterion predicts a minimum defect speed below which 
the fluid of light should not be excited substantially; this we term ``superfluid-like'' behaviour. Note also that, within this restricted range of excitation processes, there is often also  a maximum defect speed,
above which the defect is moving too fast to excite the sonic branch of Bogoliubov waves but too slow to excite the other branches, again leading to apparent low dissipation. At even higher speeds, dissipation increases again as the defect moves fast enough to excite other branches.  

In analogy with previous experiments with polariton fluids in semiconductor microcavities~\cite{Amo:NPhys2009,Amo:2011Science,Nardin:2011NatPhys}, a natural strategy to experimentally address superfluidity in our platform  would then be to look at the perturbation that is induced in the density profile by a uniformly moving defect. We also performed experiments along these lines, but had to fight serious experimental obstacles resulting from the finite size of our  photon fluid.
Its resulting rapid expansion together with the corresponding drop of its density made it difficult to extract conclusive information from the observations (see Supplemental Material~\cite{supmat}).

\begin{figure}[!]
	\includegraphics[width=1\linewidth]{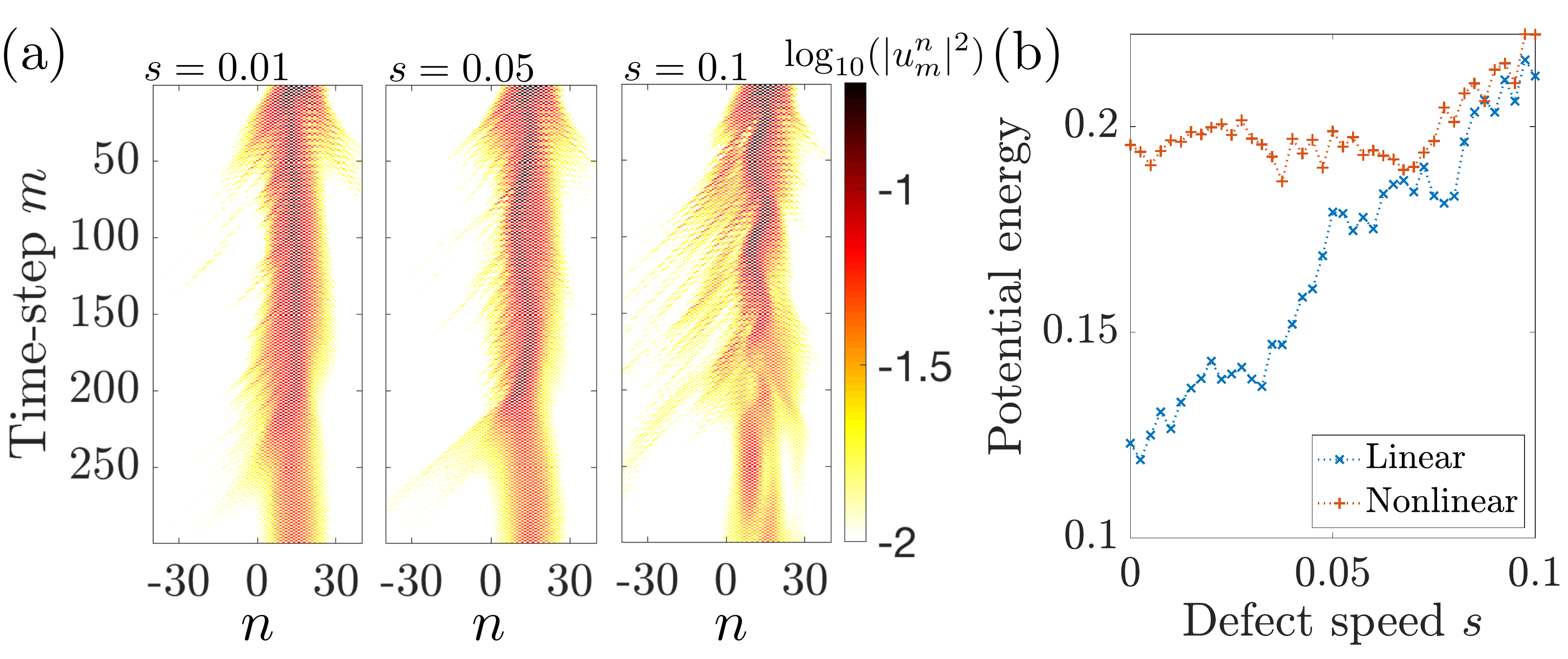}
	\caption{(a) Spatio-temporal plot of the measured intensity in the short loop for a trapped nonlinear fluid of light ($\Gamma I \approx 0.5$ in the center of the trap) excited by a sinusoidally-moving defect at various speeds $s$, applied to the optical field from $m=0$ to $m=200$. Defect and trap parameters are as given in the main text. (b) Late-time average of the potential energy~\eqref{potenergy} for experiments in the linear and nonlinear regime (nonlinear regime: power levels as in (a), linear regime: power levels 1/5 of (a)). In the linear regime, we observe that the potential energy increases steadily with the defect speeds, while in the nonlinear regime, it remains approximately constant for low $s$ and shows a marked upward kink at $s \approx 0.07$.
	}\label{sinusoidal}	
\end{figure}

To overcome this difficulty, a conceptually different scheme was adopted, so as to exploit the peculiarities of our platform. Instead of looking at the instantaneous density perturbation, our observable is the total energy that is deposited by the moving defect into the fluid during the whole excitation sequence. Related calorimetric schemes were used to detect superfluidity in atomic gases~\cite{PhysRevLett.83.2502,desbuquois2012superfluid,Weimer2015} but were never implemented in fluids of light, mostly because of the intrinsic dissipation of microcavity systems. 

More specifically, we adopted a configuration in which the optical field is excited with a defect over many time-steps and the fluid is kept in place by a confining potential. For technical reasons, the trap and defect are only applied in the short loop, i.e. $\phi_n^m=0$. The phase-shift of the short loop is the sum $\varphi_n^m = \Phi_d + \Phi_{{t}}$ of the defect contribution $\Phi_d$ given in~\eqref{eq:defectform} and the trap one, $\Phi_{{t}} = \varphi_{{t}} \left( 1- e^{- (n - n_t)^2 / \sigma_t^2} \right) $. Values  $\varphi_{{t}}=\pi / 10$,  $\sigma_t=8$ and $n_t=14$ 
of, respectively, the height, offset and half-width of the trap are chosen in order to minimize residual motion of the trapped beam, which is initialized with the Gaussian profile detailed above. 
The defect is kept in the central trapped region and is moved in space along the sinusoidal trajectory defined by $n_d(m) = 4 \sin (s m)$, and has $\sigma_n=1$ and $\varphi_d=-\pi/10$. Rather than slowly switching on and off, the defect is applied with a constant amplitude from $m=0$ until $m=200$ so to excite the fluid of light in a more significant way. 

Excitation of the optical field by the defect is investigated looking at the spatio-temporal intensity distribution shown in Fig.\ref{sinusoidal}. For all velocities, light has to adapt to the constantly changing defect potential and some light ejection can be clearly observed as light bursts propagating away from the central trapped region. But, most importantly for our superfluidity purposes, for low values of the peak defect velocity ($s=0.01$ and $s=0.05$) the field quickly returns to an almost quiet state once the defect has disappeared. Only a faster  moving defect can efficiently transfer energy to the fluid and thus permanently change its state. This heating effect is visible for $s=0.1$ as  persistent oscillations and is a clear evidence that the defect speed was large enough for superfluidity  to break down.

To make this analysis more quantitative and highlight the crucial role of superfluidity over other emission processes due to the non-inertial motion of the defect~\cite{singh2016probing}, we estimated the deposited energy by measuring the average potential energy at late times:
\begin{eqnarray}
\langle E_{\rm{pot}} \rangle = \frac{1}{j+1} \sum_{m =m_{\rm max}-j}^{m_{max}} \frac{\sum_n |u_n^m|^2\Phi_t}{\sum_n |u_n^m|^2} .\label{potenergy}
\end{eqnarray}
where $j$ corresponds to the number of time-steps after the defect is switched off; here $j=100$. This quantity is plotted in Fig.~\ref{sinusoidal}(b) for experiments in the two cases of the linear and the nonlinear regime. In the linear case, we observe that the potential energy steadily increases with the defect speed. In the nonlinear regime, we see that the potential energy starts at a higher level, 
caused by the repelling nonlinear interaction that pushes the field up the walls of the trap and
remains approximately constant for low defect speeds. A sudden threshold is visible for speeds around $s=0.07$, after which the potential energy begins to significantly increase. This behaviour can be reproduced numerically~\cite{supmat}, and confirms the presence of an effective threshold speed, above which superfluidity breaks down and friction becomes important.

{\it Conclusions -- } In this Letter we have 
reported an experimental study of superfluid light in a one-dimensional optical mesh lattice where the arrival time of pulses plays the role of a synthetic spatial dimension. The unique spatio-temporal access to the field dynamics offered by our experimental set-up was instrumental to perform the first direct measurement of the speed of sound in a fluid of light.  The conservative dynamics of our fluid of light then allowed for a quantitative measurement of the friction force felt by a moving defect and of the consequent heating effect, providing unambiguous signature of superfluidity. Taking advantage of the flexibility of the set-up in designing different geometries, on-going work is extending the investigation to fluids of light in lattices with non-trivial geometrical~\cite{wimmer2017experimental} and topological properties~\cite{bisianov,weidemann2020topological} in two~\cite{muniz} or even higher $d>3$ dimensions~\cite{RevModPhys.91.015006,ozawa2019topological}. From a more applicative perspective, the nonlinear optical mesh platform opens exciting new avenues to exploit the interplay of interference and nonlinear optical processes for manipulations of the quantum states of optical pulses in fibers~\cite{Silberhorn17,Pan17,Walmsley}.

\begin{acknowledgments}
This project was supported by German Research
Foundation (DFG) in the framework of project PE 523/14-1and by the  International Research
Training Group (IRTG) 2101.
HMP is supported by the Royal Society via grants UF160112, RGF\textbackslash{}EA\textbackslash{}180121 and RGF\textbackslash{}R1\textbackslash{}180071. IC acknowledges financial support from the European Union FET-Open grant ``MIR-BOSE'' (n. 737017), from the H2020-FETFLAG-2018-2020 project "PhoQuS" (n.820392), from the Provincia Autonoma di Trento, from the Q@TN initiative, and from Google via the quantum NISQ award. 
 \end{acknowledgments}

%

\clearpage
\widetext
\begin{center}
\textbf{\large Supplemental Material for\\ ``Superfluidity of Light and its Break-Down in Optical Mesh Lattices"}
\end{center}
\setcounter{equation}{0}
\setcounter{figure}{0}
\setcounter{table}{0}
\setcounter{page}{1}
\makeatletter
\renewcommand{\theequation}{S\arabic{equation}}
\renewcommand{\thefigure}{S\arabic{figure}}
\renewcommand{\bibnumfmt}[1]{[S#1]}
\renewcommand{\citenumfont}[1]{S#1}

\section{Basic principles of the optical mesh lattice} \label{basic}

In this section, we shall review in more detail the basic principles underlying the optical mesh lattice~\cite{schreiber2010photons,regensburger2012parity,miri2012optical}, including the justification of Eq. 1 in the main text, which governs the light evolution in our system. Further technical details about the experimental set-up are given in Supplementary Section~\ref{exptdetails}. 

\begin{figure}[b]
	\includegraphics[width=1.\linewidth]{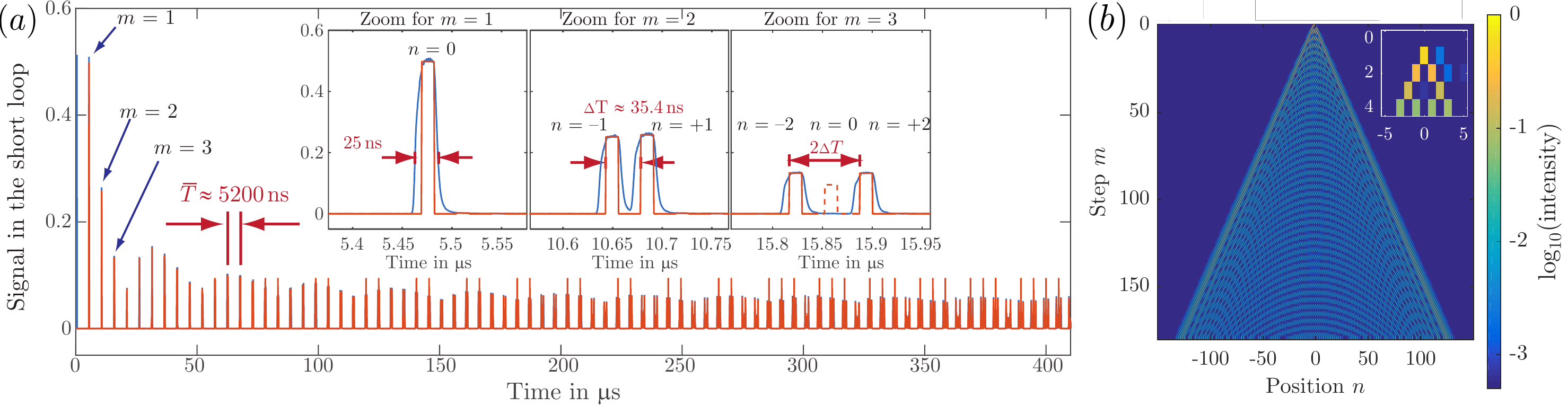}
	\caption{An example demonstrating the basic principle of an optical mesh lattice. (a) The intensity measured at a detector in the short loop for a ``Light Walk" experiment in which a single initial optical pulse is injected into a linear optical mesh lattice~\cite{regensburger2011photon, martinthesis}. Here, the blue line corresponds to the signal measured by the detector and the orange line to the average height of the pulses. As can be seen, a new group of pulses is observed after every time period $\bar{T}$, corresponding to the average time taken for a pulse to complete a round-trip in either loop.  (Here, the average loop length was roughly 1km resulting in $\bar{T}\approx 5200$ns.) Zooming in, pulses within each group are spaced out by $\Delta T$ (for this example $\Delta {T}\approx 35.4$ns) corresponding to the time-delay between a pulse travelling in the long relative to the short loop. The separation of the timescales $\bar{T}$ and $\Delta T$ allows us to label each pulse by a pair of integers ($m,n$) corresponding to, respectively, the total number of round trips and the number of extra round trips in the long rather than short loop. Note that for $(m,n)=(3,0)$ there is destructive interference between the different optical routes. (b) The pulse heights extracted from panel (a) can be replotted in terms of $m$ and $n$ to reveal the time-evolution of the light within the synthetic optical mesh lattice. In this example, the light spreads out with a characteristic ``Light Walk" distribution from a single initial pulse, with the effects of interference being clearly visible. Note that due to the intrinsic diamond connectivity of the lattice (c.f. Figure 1 in the main text), only either even or odd lattice sites are physically accessible at a given time-step. 
	}\label{overview_supmat}	
\end{figure}

As discussed in the main text, the optical mesh lattice is realised via a time-multiplexing scheme for two optical fiber loops of different length. At the start of an experimental run, an optical pulse is injected into the longer loop, and is subsequently split by the 50/50 coupler into a pulse propagating in each loop. After each round trip, each of these pulses is again split by the coupler into a further two pulses and so on. The small length difference between the loops means that the round-trip time for light travelling in the short loop, $T_2$, is shorter than that for light travelling in the long loop, $T_1$, so a pulse travelling in the short loop arrives back at the coupler before its corresponding partner in the long loop. Importantly, the average length of the loops, $\bar{L}$, is much larger than the length difference between loops, $\Delta L$, so there is a clear separation of time-scales between this relative time-delay between pulses $\Delta T = T_1-T_2$ and the average round-trip time $\bar{T}= (T_1 + T_2)/ 2$  (i.e. $\bar{T}\gg \Delta T$). Moreover, the pulse width is chosen to be smaller than the minimum pulse separation $\delta T $, so that distinct pulses do not overlap. 

As a consequence of the timescale separation, a train of distinct pulses over time is observed at a detector collecting light from one of the loops; for example, this is shown in Figure.~\ref{overview_supmat}(a) for a typical ``Light Walk" experiment in which a single initial optical pulse is injected into a linear optical mesh lattice~\cite{regensburger2011photon, martinthesis}. As can be seen, groups of pulses can be clearly identified arriving at the detector over time; we can label each of these by expressing the arrival time of any given pulse as $t= m  \bar{T} + n \Delta T/2 $, where $m$ is an integer counting the total number of round trips made by the pulse and $n$ is an integer counting how many more round trips were made in the long rather than in the short loop. It is clear that $m$ must be a positive integer, which increases by one for each group of pulses separated by $\approx \bar{T}$ going from left to right in Figure.~\ref{overview_supmat}(a). On the other hand, the integer $n$ can be either positive or negative, and can take values over the range, $-(m-1) \leq n \leq (m-1) $, for the group of pulses with a given $m$. (For example, $n=\pm (m-1)$ corresponds to light that has exclusively propagated in either the long or the short loop respectively, following the injection stage.) These observations motivate the reinterpretation of the integer $m$ as a discrete time-step, and of the integer $n$ as a discrete position index along a ``synthetic spatial dimension", leading to the effective $(1+1)$D lattice as depicted in Fig.~1(b) in the main text. 

Using this framework, we can also replot the data in Figure.~\ref{overview_supmat}(a) in terms of $(m,n)$ to obtain a result such as that shown in panel (b). Here the characteristic spreading out of the light from a single ``lattice site" can be clearly seen~\cite{regensburger2011photon}. We note that this replotting is straightforward, thanks to an unambiguous identification of $(m,n)$ for each pulse, provided that the length of the pulse sequence for a given $m$ is smaller than the round-trip time; after this, it becomes more difficult to identify individual pulses as pulses with different values of $m$ end up overlapping. This suggests a limit to the number of ``sites" along the ``synthetic spatial dimension" set by $\bar{T}/ \delta T$. As $\bar{T}$ and $\delta T$ are well-separated experimental timescales, this condition is not very restrictive as it typically allows system sizes on the order of 100 ``sites" to be considered, and can be improved further by changing the fiber-loop lengths. 

In the absence of nonlinearities or additional phase modulations, the pulse distribution at $m+1$ can be described by~\cite{regensburger2012parity,miri2012optical, martinthesis}:
\begin{eqnarray}
u_n^{m+1} &=& \frac{1}{\sqrt{2}} \left( u_{n+1}^m  + i v_{n+1}^m  \right) , \qquad \qquad
v_n^{m+1} = \frac{1}{\sqrt{2}} \left( v_{n-1}^m  + i u_{n-1}^m  \right), \label{eq:basic}
\end{eqnarray}
where $u_n^m$ and $v_n^m$ represent the (complex) pulse amplitudes incident on the beamsplitter from the short and long loops respectively. These equations simply describe the action of a 50/50 beamsplitter, where we use the standard convention that the phase of the light reflected by the beam splitter is shifted by $\pi/2$ with respect to the light transmitted. In using these equations, we are assuming that all the necessary information  is contained in these complex amplitudes, and that other effects depending on the pulse shape and finite pulse width can be neglected; these are reasonable assumptions for the parameter regimes and propagation distances considered in this experiment.

As discussed in the main text, we are interested in nonlinear phenomena which are due to the Kerr nonlinearity of the fiber, which is boosted if dispersion compensating fiber (DCF) spools form part of the long and the short loops. The nonlinear self-phase modulation of the Kerr effect then leads to a power-dependent phase-shift for the propagating light pulses, which can be expressed as $\varphi_{NL} = \gamma L_{\rm{eff}} P_{\rm{in}}$ where $P_{\rm{in}}$ is the incident power, $L_{\rm{eff}}$ is the effective length of the fibres (which also includes the effects of attenuation) and $\gamma$ is a nonlinear factor depending on the wavelength, the nonlinear refractive index and modal area of the fibre~\cite{agrawal2013nonlinear, gruner2005dispersion}. In terms of the evolution equations, this effect is captured by the addition of nonlinear phase-shifts which depend on the pulse intensities as given by~\cite{regensburger2011photon,Navarrete,wimmer2013optical} 
\begin{eqnarray}
u_n^{m+1} &=& \frac{1}{\sqrt{2}} \left( u_{n+1}^m e^{i \Gamma |u_{n+1}^m|^2} + i v_{n+1}^m e^{i \Gamma |v_{n+1}^m|^2} \right) , \qquad \qquad 
v_n^{m+1} = \frac{1}{\sqrt{2}} \left( v_{n-1}^m e^{i \Gamma |v_{v-1}^m|^2} + i u_{n-1}^m e^{i \Gamma |u_{n-1}^m|^2} \right) 
\label{eq:nonlinear}
\end{eqnarray}
where $\Gamma>0$ quantifies the effect of the nonlinearities per round trip, corresponding to a negative interaction energy in the quantum fluids language. The final ingredient to be included is that of externally-controlled phase modulators inserted in each loop, which add an additional overall phase-shift of $e^{i \varphi_n^m}$ and $e^{i \phi_n^m}$ to the short and long loops respectively. This gives~\cite{regensburger2011photon,Navarrete,wimmer2013optical} 
\begin{eqnarray}
u_n^{m+1} &=& \frac{1}{\sqrt{2}} \left( u_{n+1}^m e^{i \Gamma |u_{n+1}^m|^2} + i v_{n+1}^m e^{i \Gamma |v_{n+1}^m|^2} \right) e^{i \varphi_n^m} , \qquad v_n^{m+1} = \frac{1}{\sqrt{2}} \left( v_{n-1}^m e^{i \Gamma |v_{n-1}^m|^2} + i u_{n-1}^m e^{i \Gamma |u_{n-1}^m|^2} \right) e^{i \phi_n^m}, \label{eq:nonlinearfull}
\end{eqnarray}
which is stated as Eq. 1 in the main text.

\section{Analogy with a  Dirac Model} \label{dirac}

An analogy can be made between the above discrete evolution equations for an optical mesh lattice and a continuum  Dirac model, which exhibits similar  physics~\cite{martinthesis}. To show this, the basic linear evolution equations of the system~\eqref{eq:basic} can be written, after some algebra and relabelling, as: 
\begin{eqnarray}
\sqrt{2} (u^{m+1}_n - u_n^{m-1}) &=& (u^m_{n+1} - u^m_{n-1}) + 2 i v^m_{n}, \qquad \qquad
\sqrt{2} (v^{m+1}_n - v_n^{m-1}) = -(v^m_{n+1} - v^m_{n-1}) + 2 i u^m_{n},
\end{eqnarray}
where we recognise discretised derivatives with respect to the time-step and position on the LHS and RHS of each equation respectively. Taking the continuum limit with $\Delta t\! =\!\Delta x\! =\!1$, these equations can be re-expressed as
\begin{eqnarray}
i \frac{\partial }{ \partial t} 
\left( \begin{array}{c} \psi \\  \chi \end{array} \right) =  \frac{i \sigma_z}{\sqrt{2}}\frac{\partial }{ \partial x}  \left( \begin{array}{c} \psi \\  \chi \end{array} \right) - \sqrt{2} \sigma_x
\left( \begin{array}{c} \psi \\  \chi \end{array} \right), \label{eq:diraceq}
\end{eqnarray} 
where the two components of the vector represent the short and long loop amplitudes, and we have introduced the Pauli matrices $\sigma_z$ and $\sigma_x$. As this is a first-order equation with respect to space, this can be understood as a 1D relativistic Dirac equation. 

This reformulation into a Dirac equation provides an interesting continuum limit for our spatially and temporally discrete problem. As it usually occurs in Floquet problems~\cite{Goldman:PRX2014}, the ensuing Hamiltonian formulation is valid in the limit when the resulting dynamics is slow as compared to the temporal discreteness.
While the connection to the Dirac equation is here restricted to the linear regime, further work is in progress to extend it to the far richer nonlinear case~\cite{forthcoming}.

\section{Experimental data analysis for estimating the speed of sound} \label{exptdetails}

In this section, we detail the experimental data analysis procedures used to estimate the speed of sound displayed in Figure 2(b) of the main text. In the experiment, we rapidly turn on and off a strong stationary defect in the optical mesh lattice and observe the propagation of emitted intensity waves. The excitations are distinguished from the background optical field through the differential intensity $\Delta I= I_{{\text{pert.}}} -I_{{\text{unpert.}}}$, where $I_{{\text{pert.}}}$ ($I_{{\text{unpert.}}}$) corresponds to the local intensity for the perturbed (unperturbed) evolution, as shown in Figure 2(a) of the main text. 
Turning on and off the defect will excite the Bogoliubov dispersion [see Fig. 1 in the main text] predominately around $k=0$ and $\theta=0$. Within this region, the low-momenta sound waves are predicted to have the lowest group velocity, and so we associate the innermost minima in excitation evolution [see Fig. 2(a) in the main text] to them. In so doing, we assume that states which have an even lower group velocity (e.g.  in the second branch or at high momentum) are very off-resonant, and do not play a significant role in the excitation propagation observed here. This will be investigated in depth in an upcoming theoretical work~\cite{forthcoming}. 

\begin{figure}[t!]
	\includegraphics[width=1\linewidth]{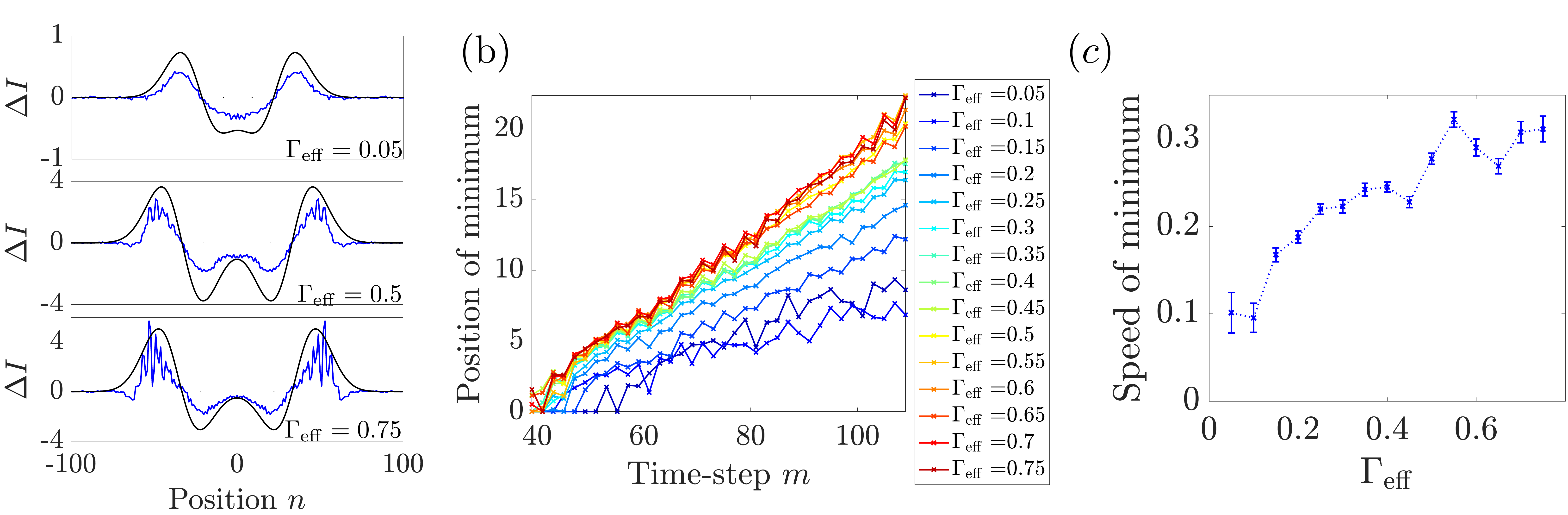}
	\caption{(a) Examples of cuts of the differential intensity at the final time-step $m=110$ for different effective initial nonlinearities. Blue line represents the experimental data and black line is the fitting function~\eqref{eq:fit}. (b) The extracted position of the innermost minimum over time from the fits in panel (a) for a range of different nonlinearities. (c) The estimated speed of the minimum as a function of the effective initial intensity as calculated from panel (b) by applying a linear fitting function~\eqref{eq:lin} over the last 30 time-steps. Errorbars show the standard error in the fit coefficient $a_2$. }\label{fit}	
\end{figure}

Identifying the innermost minima as sound waves, we then wish to extract their positions as a function of time, in order to calculate the speed of propagation. To achieve this, we fit the data at each time-step with a function: 
\begin{eqnarray}
f_1(n) =( a_1 + b_1(n-c_1))e^{-(n-c_1)^2/d_1^2} + ( a_1 - b_1(n+c_1))e^{-(n+c_1)^2/d_1^2} \label{eq:fit}
\end{eqnarray}
where $a_1, b_1, c_1, d_1$ are all fitting parameters. This function is chosen as it consistently reproduces the positions of the largest features of the perturbation pattern, namely the two inner minima and two outer maxima, although without necessarily capturing the overall amplitude of these peaks. Example results from this fitting procedure are shown in Fig.~\ref{fit}(a). The fitting works best for high effective nonlinearities and for late times, when the minima and maxima are most clearly developed. Note that we have symmetrized the experimental data with respect to $n=0$, to facilitate the fitting procedure.  Experimentally, asymmetries can arise from small imbalances in the gain and loss between the two loops or from errors in the initial state preparation, leading to an underlying overall drift of the optical field.   

After fitting the data, we numerically extract the positions of the inner-most minima, as shown in Fig.~\ref{fit}(b). As can be seen, the position increases over time corresponding to the excitation propagating outwards from the center. There are also additional small oscillations in the position, the periodicity of which coincides with the periodicity of the optical mesh lattice. These may therefore be an artefact of fitting to alternating discrete sets of points at alternating time-steps [see Fig.1(b) in the main text]. In any case, we average over these by fitting the minima positions over the last 30 time-steps with the linear function:
\begin{eqnarray}
f_2 (m) = a_2 m + b_2 \label{eq:lin}
\end{eqnarray}
where $a_2$ and $b_2$ are fitting parameters. We can then associate the value of $a_2$ with the speed, $v_M$, as is plotted in Fig.~\ref{fit}(c) and in Fig.~2(b) in the main text. Errorbars shown indicate the standard error in the fit coefficient $a_2$. We have also checked that other methods of extracting the positions of the minima, i.e. by using other fitting functions, are in reasonable agreement with the above approach.

\begin{figure}[t]
	\includegraphics[width=1\linewidth]{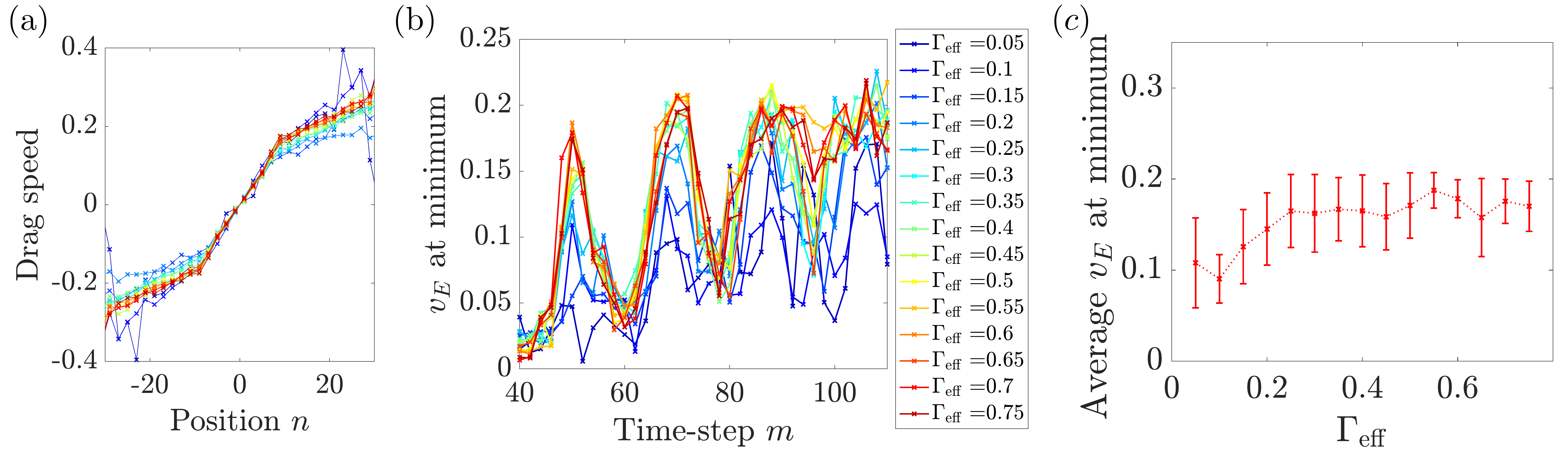}
	\caption{(a) Example of estimated drag speed using Eq.~\eqref{eq:num} for a range of effective initial nonlinearities as indexed by the legend of panel (b). The data is shown for $m=90$ with $\Delta m =4$. (b) The corresponding estimated drag speed at the position of the minimum extracted from Fig.~\ref{fit}(b), as a function of the time-step for different effective initial nonlinearities. (c) Time-average of the drag speed shown in panel (b) over the last 30 time-steps of evolution. Points are connected by lines as a guide to the eye. Errorbars show the standard deviation in $v_E$ over the last 30 time-steps.}\label{drag}	
\end{figure}

As discussed in the main text, we also need to take into account the underlying expansion of the light field over the propagation time, as this will drag the perturbations outwards and increase their measured speed. To estimate this, we apply a version of the local 1D continuity equation with respect to a given position $A$:
\begin{eqnarray}
\frac{d I_A}{  d t} = -v_E\frac{I (A)}{2}  + \beta I_A \label{eq:cont}
\end{eqnarray}
where $v_E$ is the drag speed, $\beta$ is an overall amplification factor and $I_A = \sum_{n \in \mathcal{A}} I(n)$ 
is the integrated local intensity, $I (n) = |u_n|^2 + |v_n|^2$ in the absence of the defect. The summation domain $\mathcal{A}$ is chosen such that either $-\infty \!\leq\! n\!\leq\!  A$ when $A>0$ or $A\! \leq\! n\!\leq\!  \infty$ when $A<0$, to ensure that the sum always covers at least half of the light field to minimize the effects of noise. Physically, Eq.~\eqref{eq:cont} equates the rate of change of the intensity within the chosen region to the amount of light flowing in/out and the net gain within that region. Note the factor of $1/2$ in the first term on the RHS takes into account that the lattice spacing is $2$ in the optical mesh lattice (see Fig.~1(b) in the main text), and hence the local density of light is given by ${I (A)}/{2}$. In the optical mesh lattice, time is discretized and the above continuity equation can be approximated as:
\begin{eqnarray}
\frac{ I_A (m) - I_A (m- \Delta m) }{  \Delta m }=-v_E\frac{I (A)}{2} + \beta I_A (m) \label{eq:num}
\end{eqnarray}
where we explicitly include the dependence on the time-step. To determine the overall amplification $\beta$, we apply this equation to the summed intensity over the entire system:
\begin{eqnarray}
\frac{ I_\infty (m) - I_\infty (m- \Delta m) }{  \Delta m }= \beta I_\infty (m).
\end{eqnarray}
In the experiments we used Erbium-doped amplifiers together with acousto-optical modulators  to compensate for losses, thus realizing a conservative situation. Still the values of $\beta$ are found to be non-zero and slightly positive, corresponding to small net gain in each roundtrip. The values of $\beta$ are used to calculate $v_E$ from~\eqref{eq:num}; this numerically-extracted drag speed is shown in Fig.~\ref{drag}(a), for time-step $m=90$ with $\Delta m =4$ [colors indicate effective nonlinearity as indicated in the legend of panel (b)].  Note that we have again symmetrized the underlying experimental data about $n=0$, to be consistent in the data analysis. 

In terms of validity, this method for estimating $v_E$ is expected to break down when the intensity is too low, as it will then be more sensitive to experimental noise and discretization errors. This is observed numerically by going to larger positions $n$, and is likely responsible for the sharp variations at the lowest nonlinearity shown in Fig.~\ref{drag}(a). We note that this method also assumes that the amplification $\beta$ is uniform within a single round trip, a condition which is well realized by Erbium-doped amplifiers due to their slow response time.

To estimate the effect of the expansion on the motion of the perturbation, we track the drag speed at the position of the innermost minima from the perturbed evolution, i.e. we set $A$ equal to the extracted position of the minimum from Fig.~\ref{fit}(b). Note that the latter is obtained as the minimum of a fitting function and so can take continuous values, while the former is constrained to take only certain discrete values corresponding to the available lattice sites of the optical mesh lattice [see Fig. 1(b) in the main text], and so this is an approximation. We also point out that the drag speed is calculated from the unperturbed evolution in order to provide a measure of the underlying expansion; in so doing, we assume that the excitations are not strong enough to significantly alter the effective drag speed. 

The resulting drag speed at the position of the innermost minimum is plotted in Fig.~\ref{drag}(b). Note that as the drag speed is symmetric about $n=0$, either the positive or negative minimum position can be taken. As can be seen, the behaviour of the drag speed appears to be reasonably similar for high nonlinearities as the drag speed does not vary significantly over the respective positions of the minima [see Fig.~\ref{drag}(a)]. There are also large oscillations in the drag speed for all nonlinearity strengths, which occur approximately every twenty time-steps and which likely correspond to small oscillations that we observe in the intensity of the unperturbed evolution. As they are not associated with the excitations of interest, we average over the oscillations during the last 30 time-steps to obtain an estimate for the average expansion speed over the period in which the speed of the minima is extracted. This averaged speed is plotted in Fig.~\ref{drag}(b) and in Fig.~2(b) in the main text. Errorbars show the standard deviation in $v_E$ over the last 30 time-steps. The speed of sound is then calculated as $v_S= v_M - v_E$ as shown in Fig.~2(b) in the main text. Errorbars in $v_S$ are found by propagating the estimated errors in $v_M$ and $v_E$ as shown in Fig.~\ref{fit}(c) and Fig.~\ref{drag}(c) respectively.

\begin{figure}[t!]
	\includegraphics[width=1\linewidth]{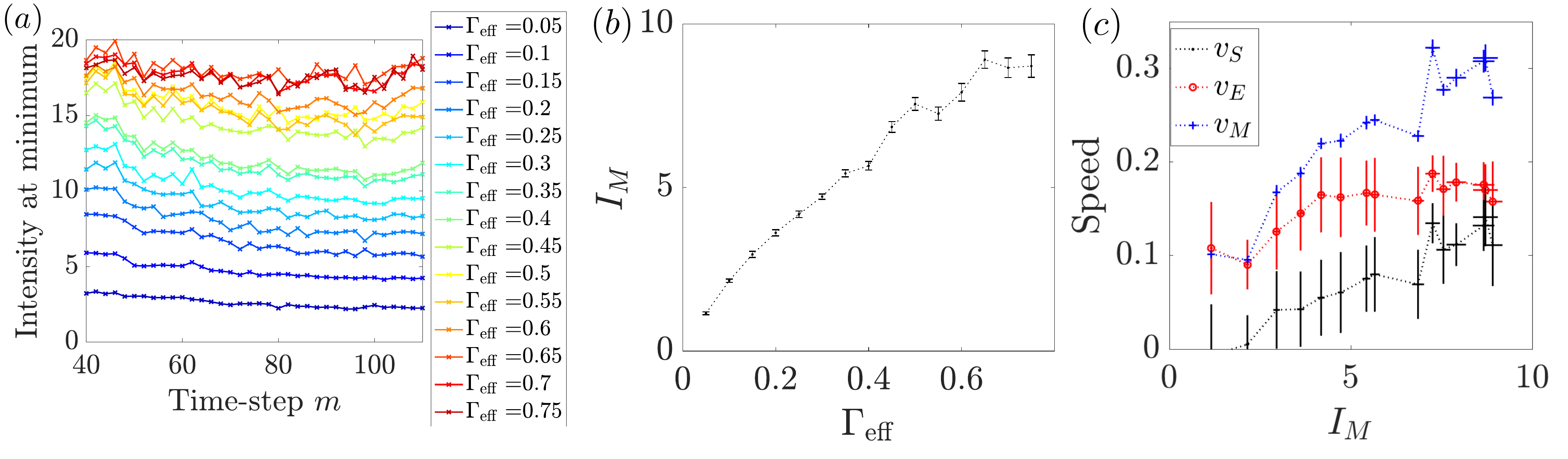}
	\caption{(a) The unperturbed total local intensity (summed over both loops in arbitrary units) at the position which corresponds to the minimum in the perturbed experiment. (b) The corresponding average local intensity, $I_M$, in each loop, which has been calculated over the last 30 time-steps from panel $(a)$ and is plotted versus $\Gamma_{\rm{eff}}$.  (Error-bars show the standard deviation in the intensity over the last 30 time-steps.)
	(c) The estimated speed of the minimum ($v_M$), drag speed ($v_E$) and speed of sound ($v_s$) replotted as a function of $I_M$, showing the same qualitative trends as in Figure 2 in the main text. Points are connected by dotted lines as a guide to the eye. Vertical errorbars show the standard deviation in each quantity as shown previously. Horizontal errorbars show the standard deviation in $I_M$ over the last 30 time-steps.}\label{im}	
\end{figure}

Finally, note that Fig.~\ref{fit}(c), Fig.~\ref{drag}(c) and Fig.~2(b) in the main text have been plotted as a function of the initial effective nonlinearity $\Gamma_{\rm{eff}}$, as defined from the initial peak intensity of the light field. In practice, the intensity and hence the effective nonlinearity and local speed of sound can change over time as the light field expands. This can be seen in Figure~\ref{im}(a), where we plot the total unperturbed intensity (summed over both short and long loops in arbitrary units) at the position which corresponds to that of the minimum in the perturbed experiment. Assuming that the defect only creates weak perturbations, this total local intensity is related to the local nonlinearity and determines the local sound speed.

However, in the experiment the intensity level is partially stabilised by gain from the Erbium-doped amplifiers; in the absence of gain, the intensity would drop more dramatically as the narrow light field expands, as shown in theoretical simulations in Fig.~\ref{st2}. Notwithstanding these complications, it can be seen in Figure~\ref{im}(a) that a higher initial peak intensity (i.e. a higher effective nonlinearity $\Gamma_{\rm{eff}}$) leads to a higher local intensity at the position where the minimum would be. The relationship between these two quantities can be even more clearly seen in Figure~\ref{im}(b), where we plot $I_M$ versus $\Gamma_{\rm{eff}}$; the former is defined as the local averaged intensity in each loop and is calculated from the total summed intensity shown in panel (a) by averaging over the last 30 time-steps and dividing by two to account for the two loops.  Error-bars show the standard deviation in the intensity over the same time period. 

Using this, we can re-plot the data in Fig.~2(b) in the main text as a function of $I_M$, instead of $\Gamma_{\rm{eff}}$, as shown in Figure~\ref{im}(c). Here, the vertical error bars are the same as in the previous figures, while the horizontal errorbars indicate the standard deviation in the averaged intensity as in Figure~\ref{im}(b). As can be seen, the re-plotting of the data reorders and shifts a few points by small amounts, but does not affect the the main qualitative feature of interest, i.e. that the apparent speed of sound increases from around zero with increasing nonlinearity. Our conclusions therefore remain the same regardless of whether we consider the initial peak intensity or the final (unperturbed) local intensity as an experimental estimate of the nonlinearity present in the system.

\section{Numerical simulations for estimating the speed of sound}

\begin{figure}[t!]
	\includegraphics[width=1\linewidth]{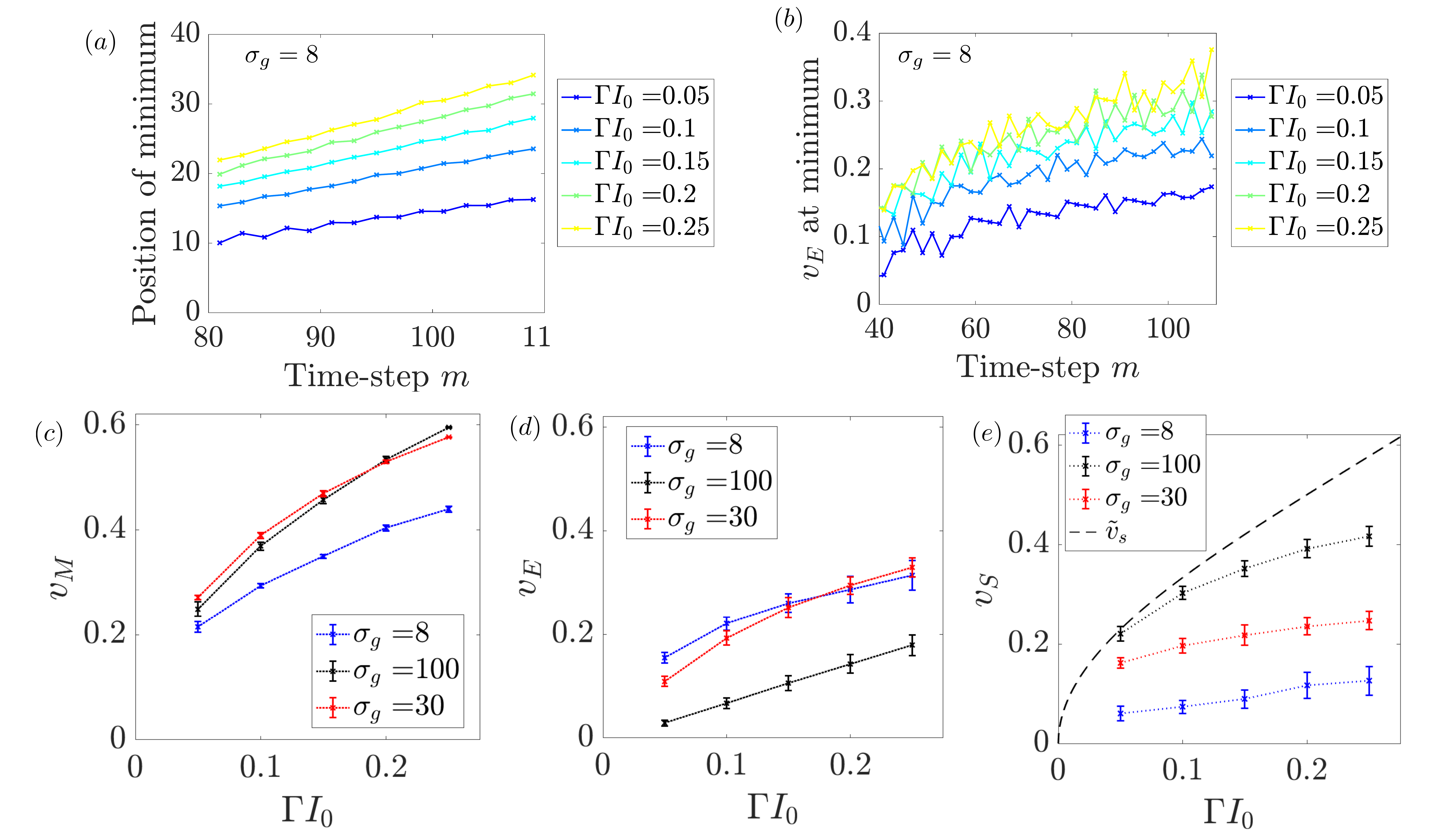}
	\caption{ (a) The fitted position of the minimum in numerical simulations with $\sigma_g=8$ and other parameters as given in the text. (b) The extracted drag speed $v_E$ for the same numerical simulations as shown in (a). (c)-(e) The estimated speed of the minimum ($v_M$), drag speed ($v_E$) and speed of sound ($v_s$) plotted as a function of $\Gamma I_0$ for three different initial light-field widths. Points are connected by dotted lines as a guide to the eye. Vertical errorbars show the standard deviation in each quantity, calculated in the same way as for the experimental data shown above. As can be seen, the results significantly underestimate the expected speed of sound $\tilde{v}_s$ (black dashed line, Eq.~3 in the main text), with the deviation increasing for decreasing initial light field width $\sigma_g$. }\label{st1}	
\end{figure}

To further verify our approach, we have applied our experimental data analysis procedures to numerical simulations, so that a direct comparison to the analytical speed of sound can also be made. As in the previous section, our first step is to extract the position of the minimum over time by fitting our numerical simulations; an example set of numerical results are shown in Figure~\ref{st1}(a) for five different values of $\Gamma I_0$ with an initial light-field width set by $\sigma_g=8$, and other parameters chosen as those in the experiment, i.e. $\varphi_d = \pi /10$, $ \sigma_n=1$, $\sigma_m=2$, $m_d=20$, $m_{\rm{max}}=110$. Note that numerically, we fix $I_0 = 1$ and then vary $\Gamma$ in order to change the effective nonlinearity parameter, $\Gamma I_0$, while keeping the initial light field the same. As with the experimental data, we find that the fitting procedure (to extract the position of the minimum) works best for high effective nonlinearities and for late times when the minima and maxima are most clearly developed. 

The second step is then to calculate the drag speed $v_E$ due to the expansion of the light field at the fitted position of the minimum, as described in the previous section; the result of this is shown for the same example set of simulations in Figure~\ref{st1}(b). (Note that $\beta=0$ as gain is not included in the numerics for simplicity.) In contrast to Fig.~\ref{drag}(b), we do not observe any large oscillations in $v_E$, supporting our conclusion that these were not associated with the excitations of interest. 

\begin{figure}[t!]
	\includegraphics[width=1\linewidth]{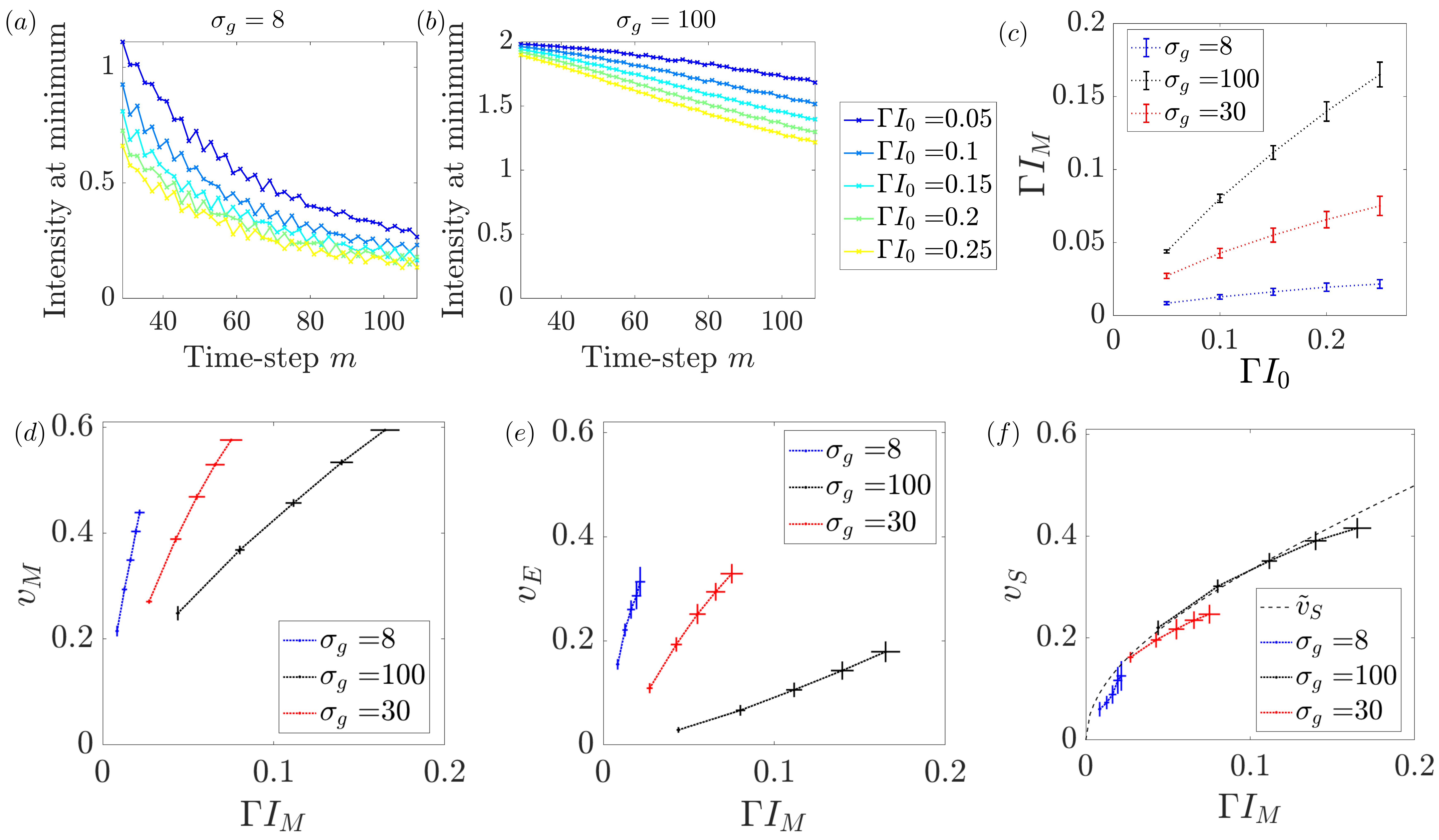}
	\caption{The numerical unperturbed local intensity (summed over both loops) at the position which corresponds to the minimum in the perturbed numerical run for (a) $\sigma_g=8$ and (b) $\sigma_g=100$. Note that here the inital peak intensity is fixed, while the parameter $\Gamma$ is changed. (c) The corresponding local nonlinearity in each loop, $\Gamma I_M$, averaged over the last 30 time-steps and plotted versus $\Gamma I_0$.  (Error-bars show the standard deviation in $\Gamma I_M$ over the last 30 time-steps.) Note that in the experiment it was easier to change the intensity between runs (as $\Gamma$ is fixed by the optical nonlinearity of the fibers), while, numerically, we chose to keep the initial intensity constant and then to change the value of $\Gamma$. This means that the ordering of the colours from top to bottom is reversed in panels (a) and (b) as compared to Figure~\ref{im}(a)). However, this difference is not important as the relevant nonlinearity parameter is the combination of both intensity and $\Gamma$, and in panel (c) we observe similar positive correlations between the initial and local nonlinearities as for the experiment in Fig.~\ref{im}(b). 
	(d)-(f) The estimated speed of the minimum ($v_M$), drag speed ($v_E$) and speed of sound ($v_s$) replotted as a function of $\Gamma I_M$. Points are connected by dotted lines as a guide to the eye. Vertical errorbars show the standard deviation in each quantity as shown previously. Horizontal errorbars show the standard deviation in $\Gamma I_M$ over the last 30 time-steps. In panel (f) all curves collapse towards the the analytical local speed of sound $\tilde{v}_s$ (black dashed line), which has been calculated by replacing $\Gamma I_0$ with $\Gamma I_M$ in Eq.~3 in the main text.}\label{st2}	
\end{figure}

In the final step, we extract the averaged values for $v_M$, $v_E$ and $v_s$ over the last 30 time-steps from the numerical results, as plotted in Figure~\ref{st1}(c)-(e). (Errorbars represent the standard deviation from the fitting procedures as detailed in the previous section.) As can be seen from panel (e), we observe a large deviation between the calculated speed of sound for $\sigma_g=8$ and the analytical speed of sound, $\tilde{v}_s$ (black dashed line, Eq.~3 in the main text). We have also repeated our simulations for two different initial light-field widths (set by $\sigma_g=20,100$), and we observe that the agreement between the numerics and the analytics appears to improve for increasingly wide light fields. However, even for the very wide initial light field ($\sigma_g=100$), there appears to still be significant deviation between the numerics and the analytics at larger nonlinearities. As we now show, both the apparent deviation between the analytics and the numerics as well as the strong dependence of the numerics on the inital light-field width can be understood by considering what happens to the {\it local} intensity experienced by the sound waves in these simulations. To see this, we plot the total unperturbed intensity (summed over both short and long loops) at the position corresponding to that of the minimum in Figure~\ref{st2}(a) and (b) for $\sigma_g=8$ and $\sigma_g=100$ respectively. (Note that as we have fixed $I_0 = 1$, the total initial peak intensity summed over both loops is equal to $2I_0 =2$ in all cases.) As can be seen, for the narrow light-field ($\sigma_g=8$), the total local intensity at the minimum drops very quickly over time, while for the wide light-field ($\sigma_g=100$), the total local intensity remains more stable (although still decreasing) as there is a wide central region within which the intensity remains close to $2 I_0$ over these timescales. We can also see that a higher nonlinearity parameter $\Gamma$ causes the intensity to drop more quickly as this physically corresponds to having stronger repulsive interactions. 

As the speed of sound depends on the local intensity, these intensity variations imply that the numerical results from Figure~\ref{st1}(c)-(e) should be replotted as a function of the local intensity rather than of the initial intensity. To do this, we first plot $\Gamma I_M$ as a function of $\Gamma I_0$ in Figure~\ref{st2}(c), where $I_M$ is the averaged local intensity in each loop. The latter is calculated by averaging the total intensity over both loops (e.g. from Figure~\ref{st2}(a) and (b)) over the last 30 time-steps and then dividing by two to find the average local intensity in each of the two loops. Here, errorbars show the standard deviation from averaging the intensity. Note that time-averaging the local intensity introduces a further approximation in our analysis as the intensity and hence the non-linearity does drop over 30 time-steps (see e.g. Figure~\ref{st2}(a) and (b)); this will be studied in more detail in further work~\cite{forthcoming}. Nevertheless, we can use Figure~\ref{st2}(c) to replot the data from Figure~\ref{st1}(c)-(e) in terms of the local nonlinearity parameter, $\Gamma I_M$, as shown in  Figure~\ref{st2}(d)-(f). Crucially, the extracted speed of sound $v_s = v_M - v_E$ for different initial widths now collapse towards the analytical local speed of sound $\tilde{v}_s$ which is calculated by replacing $\Gamma I_0$ with $\Gamma I_M$ in Eq.~3 in the main text. The remaining deviations from the expected speed of sound are due to the approximations made in our analysis, and to the choice of parameters~\cite{forthcoming}. This demonstrates that by taking into account the behaviour of the local intensity, we can remove the apparent dependence on the initial light-field width and recover the expected analytical result, verifying our data analysis protocols. 
 
These issues turn out to be much less important in the experiment, where the local intensity is stabilised by gain (c.f. Figure~\ref{im}(a)), and so does not drop as dramatically as in these numerical simulations. As a consequence of this, we expect that the deviations from the expected speed of sound and the dependence on the initial light-field width will be less significant in the experiment than in the numerics. Moreover, for the experiment, we were only concerned with overall qualitative trends, as we were not able to extract the absolute value of the nonlinearity parameter. In both the numerics [Figs.~\ref{st1} and~\ref{st2}] and the experiment [see the main text and Fig.~\ref{im}], these overall trends are unaffected by the replotting of the data in terms of the local nonlinearity, showing that this replotting is not essential for the qualitative conclusions presented in the main text. With this in mind, we have therefore chosen to focus on the experimental data plotted as a function of the initial effective nonlinearity in the main text as this is both conceptually simpler and requires fewer data processing steps and approximations.

\begin{figure}[!]
 	\includegraphics[width=0.99\linewidth]{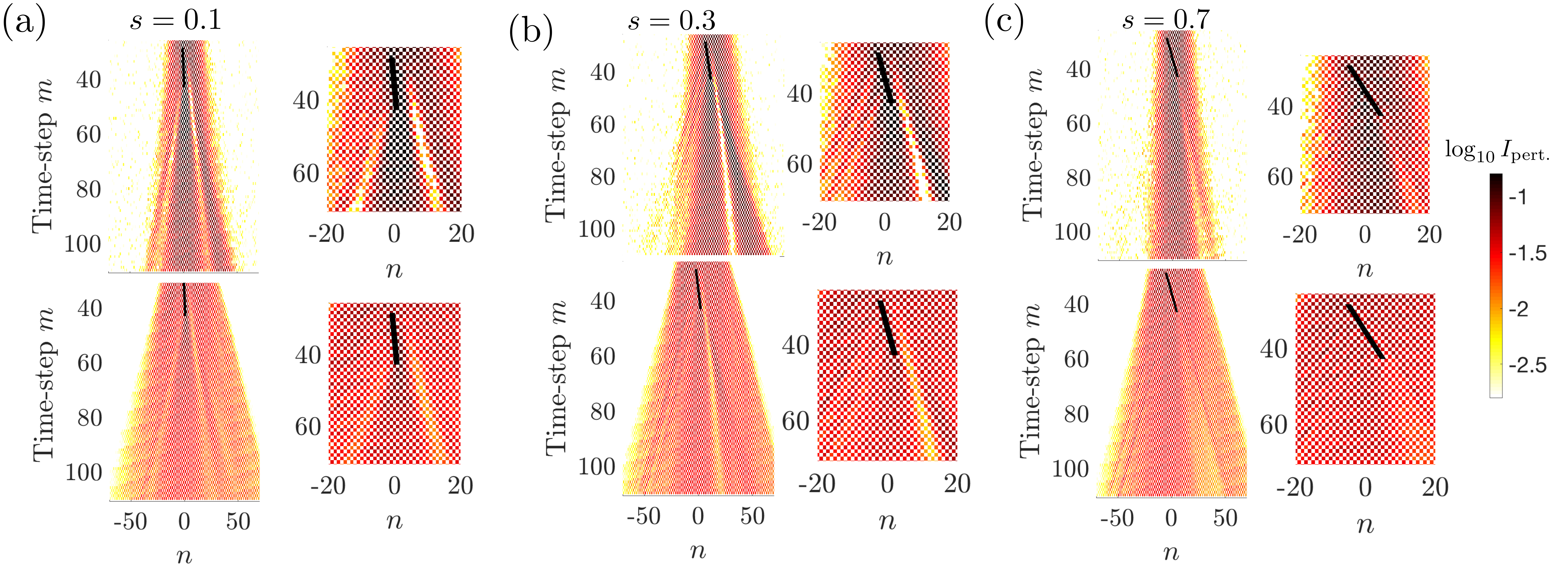}
 	\caption{
 	The perturbed intensity, $I_{{\text{pert.}}}$, as a moving defect is switched on and off with various defect speeds. The initial light field is a Gaussian beam prepared in the linear ({\it Top Row}) and nonlinear ({\it Bottom Row}) regime for defect speed: (a) $s=0.1$, (b) $s=0.3$ and (c) $s=0.7$. The initial peak light intensity in the nonlinear regime is approximately 14 times higher than in the linear regime. 
 	Within each panel, we plot the full intensity ({\it left}) and a zoom-in around the defect ({\it right}). Black solid lines represent the defect; these lines are drawn on top of the data to indicate the FWHM of the Gaussian defect profile with respect to both time-step and position. Parameters of the defect and initial light field are as stated in the text. In the linear regime, we observe strong excitation of the optical field for low defect speeds, which decreases for large defect speeds. In the nonlinear case, we observe that the excitation amplitude appears to increase at intermediate defect speeds around $s=0.3$, before again falling off at larger defect speeds.}\label{full}	 \end{figure}

\section{Defect moving uniformly across an optical beam}

As a first attempt to investigate the superfluid-like behaviour of light in the nonlinear optical mesh lattice, we move a defect uniformly across the optical light field at different speeds. In Fig.~\ref{full}, the panels in the top (bottom) row show the resulting perturbed intensity, $I_{{\text{pert.}}}$, for a linear (nonlinear) experiment.	The defect profile $\varphi_d$, is imprinted via the phase shift in the short loop as
\begin{eqnarray}
 \varphi_{n}^m = \phi_{n}^m =\Phi_d \equiv \varphi_d e^{-(n - n_d)^2 /  \sigma_n^2} e^{-(m - m_d)^2 / \sigma_m^2} \label{eq:defectform1}
\end{eqnarray}
with $n_d= s(m-m_d)$ and $s$ being the constant speed. In the experiment, we take $\varphi_d =- \pi /10$, $\sigma_n=1$, $m_d=35$ and  $\sigma_m=10$ to ensure that the defect does not leave the central region of the light field, which is again initialised with the Gaussian profile described in the main text. As can be seen, this defect profile influences the light field in both regimes even at low speeds as the field adapts to the time-dependent potential and excitation of the Bogoliubov branches may take place.

 In the linear regime, the strength of the excitation does not significantly vary with defect speed, until it falls off at high speeds because of the upper critical velocity effect~\cite{forthcoming}. We note that at high speeds, the defect quickly leaves the central high intensity region making it difficult to directly compare results. In the nonlinear regime, we observe an enhancement of excitations at intermediate speeds (around $s=0.3$), with a drop-off at lower and higher speeds. While this qualitative behaviour is in keeping with there effectively being a speed range over which the defect most efficiently excites the optical field, this set of data are not clear enough to perform a quantitative analysis of the critical velocities. As it is discussed in the main text, quantitative information on the critical velocity can be extracted by implementing a different, sinusoidal, motion of the defect.

\section{Defect moving sinusoidally in a Gaussian trapped optical beam}

In this section, we numerically simulate a defect moving sinusoidally in a Gaussian trap to show that we recover the qualitative features of the observed experimental behaviour shown in Fig. 4 in the main text. This is important to further confirm that the observed excitation is dominated by breakdown of superfluidity at high speeds, rather than to non-inertial effects due to the sinusoidal (instead of linear) motion of the defect.

The defect profile $\varphi_d$, is imprinted via the phase shift in the short loop as
\begin{eqnarray}
  \varphi_{n}  = \Phi_d\equiv\varphi_d e^{-(n - n_d)^2 /  \sigma_n^2}  \label{eq:defectform}
\end{eqnarray}
where we choose $n_m = 4 \sin (sm)$, with $n_d=1$ and $\varphi_d=-\pi/10$. The defect is applied with this constant amplitude from $m=0$ until $m=200$. The Gaussian trap is also imposed through the phase-shift of the short loop: $\varphi_n^m = \Phi_d + \Phi_{{t}}$ where $\Phi_{{t}} = \varphi_{{t}} \left( 1- e^{- (n - n_t)^2 / \sigma_t^2} \right) $, with $\varphi_{{t}}$, $n_t$ and $\sigma_t$ being, respectively, the height, offset and width of the trap. Here, we take $\varphi_{{t}}=\pi / 10$,  $\sigma_t=8$ and $n_t=0$. The optical field is initially prepared, as in the experiment, with a Gaussian profile corresponding to an equal amplitude in each loop at $m\!=\!0$: $u_n^0 \!= \!v_n^0\!=\! \sqrt{I_0} \rm{mod} (n,2) e^{- n ^2 / \sigma_g^2}$, where we take $\sigma_g= 8$ as the Gaussian width. Numerically, we take $\Gamma=0.5$ and then change $I_0$ in order to change the effective nonlinearity $\Gamma I_0$.  We note that despite choosing experimentally-realistic parameters as much as possible, the following results are not intended to be quantitatively compared to experiments, as is it not known what the absolute light intensity is experimentally nor what the appropriate gain or loss per round-trip is for each level of nonlinearity (which here, for simplicity, is set to zero).   

\begin{figure}[!]
	\includegraphics[width=1\linewidth]{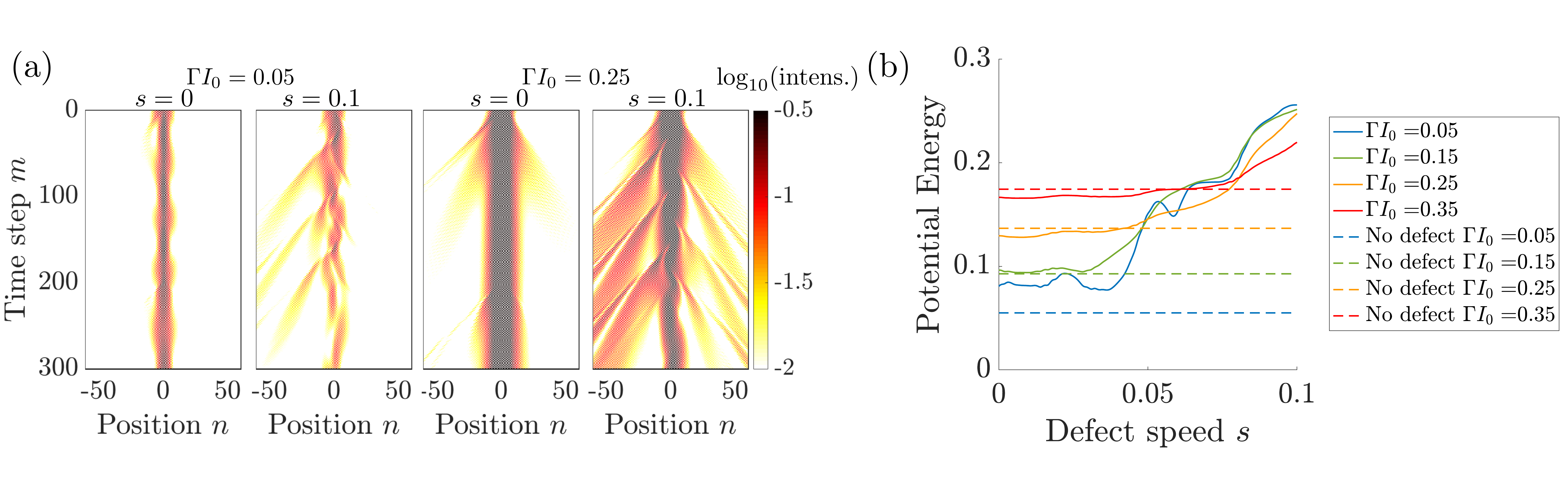}
	\caption{(a) Numerical intensity in the short loop for a trapped nonlinear light field excited by a sinusoidally-moving defect at various speeds $s$ and for different nonlinearities. All parameters are given in the text. (b) The corresponding average potential energy~\eqref{potenergy}. Solid (dashed) lines indicate the results in the presence (absence) of a moving defect. }\label{sin}
\end{figure}

The intensity in the short loop over the propagation time is shown in Fig.~\ref{sin}(a) for two different values of the nonlinearity. As can be seen, there is some residual excitation of the optical field even when $s=0$, corresponding to a stationary defect, due to the switching on and off of the defect potential. When the defect is moved, it can excite the optical field leading to the emission of light outside of the trap. As stated in the main text, we try to quantify this emission by estimating the average potential energy after the defect has been switched off:
\begin{eqnarray}
\langle E_{\rm{pot}} \rangle = \frac{1}{j+1} \sum_{m =m_{\rm max}-j}^{m_{max}} \frac{\sum_n |u_n^m|^2\Phi_t}{\sum_n |u_n^m|^2} ,\label{potenergy}
\end{eqnarray}
where $j$ corresponds to the number of time-steps after the defect is switched off; here $j=100$. This is plotted in Fig.~\ref{sin} for a range of different nonlinearities. We have also plotted for comparison the numerical potential energy for each case without a defect $\varphi_d=0$; as can be seen, this increases steadily with the initial intensity and hence the nonlinearity 
because the field is pushed away from the trap center due to the defocusing nonlinearity.
Introducing the defect, we observe that at very low nonlinearities $\Gamma I_0=0.05$, the potential energy oscillates, probably due to the low intensity level of the optical beam which is significantly perturbed by the defect (see Fig.~\ref{sin} (a)). At larger nonlinearities, the behaviour is much smoother, and we clearly observe that the potential energy initially remains close to that without the defect, before beginning to rise after a threshold speed which increases with increasing nonlinearity. This confirms the  interpretation of the qualitative features observed experimentally in Fig. 4(b) in the main text in terms of the Landau criterion for superfluidity.

\section{Experimental Details} \label{exptdetails}

\begin{figure*}[!]
	\includegraphics[width=0.7\linewidth]{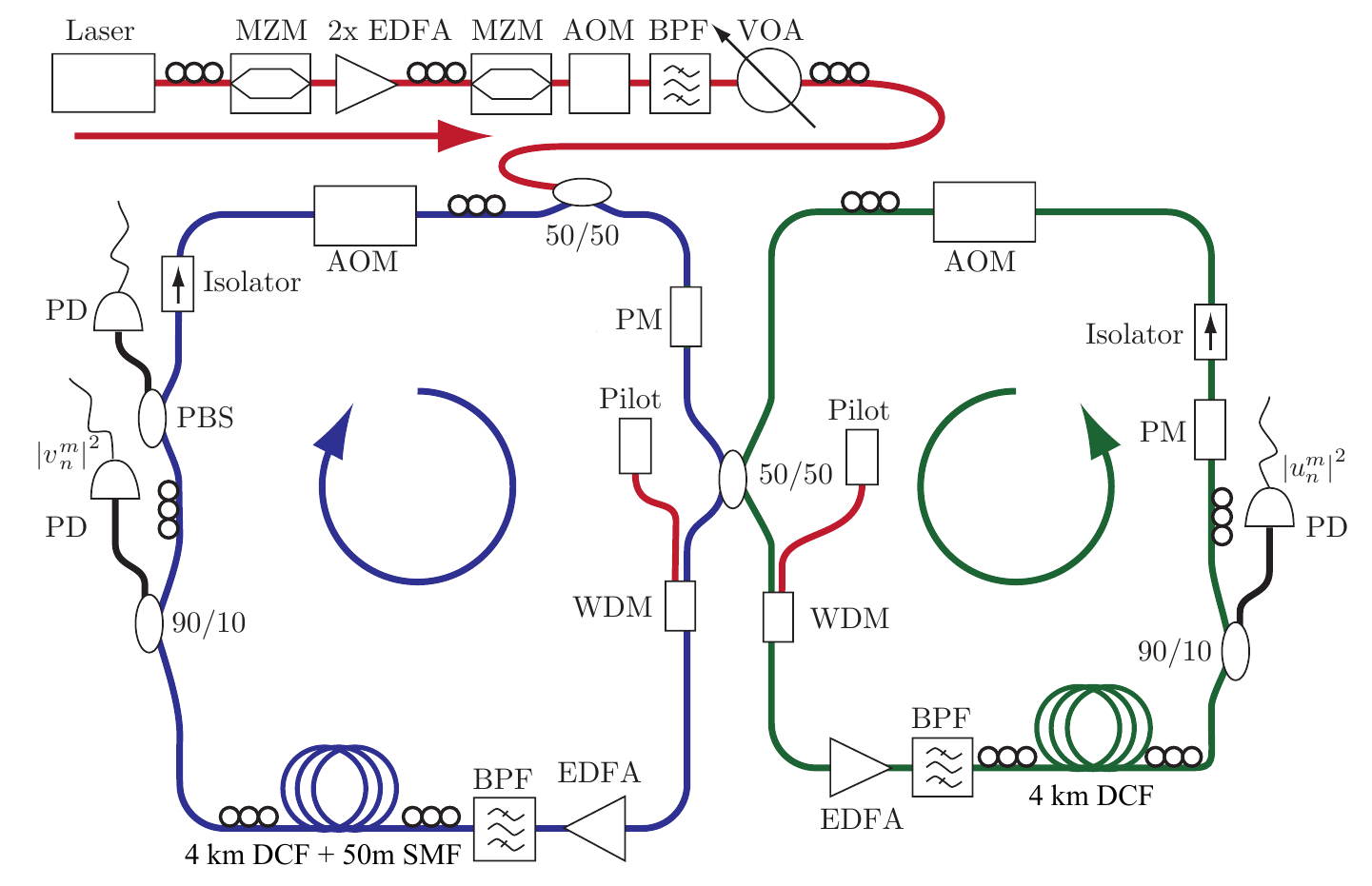}
	\caption{Experimental set-up. 
	}\label{experiment}	
\end{figure*}
 At the beginning of each measurement, a pulse ($\lambda_{signal} = 1555$nm) is shaped by the signal generator unit. To this end the signal of a DFB laser diode is cut into pulses
of 50 ns width by a Mach-Zehnder Modulator (MZM). Two subsequent erbium-doped
fibre amplifiers (EDFA) increase the peak power of the pulses. In the experiment, power levels
of up to 110mW are reached. Afterwards
another MZM suppresses the background between two pulses for increasing the contrast
between the coherent pulses and the noisy background. An acousto-optical modulator (AOM)
is needed as a gate for blocking further pulses during the actual measurement. For
cleaning the spectrum of the generated pulses, a tunable bandpass filter is placed after the
AOM. Finally, for manually varying the height of the pulses, a variable optical attenuator
(VOA) is used. The AOM can also be driven in an analogue mode allowing for an automatized
sweep of the initial power and for a fine-tuning of the initial peak intensity. Due to integrated
polarizers in the MZM, a polarization controller is placed in front of each, and additionally a
third one is placed directly before the 50/50 input coupler to control the initial polarization
of the pulses.
After insertion  into
the longer  loop, the pulse splits up at the central 50/50 coupler into two smaller
pulses. In each loop, the pulses pass first a wavelength division multiplexing (WDM) coupler,
which adds a pilot signal operating at a shorter wavelength $\lambda_{pilot} = 1536$nm. The pilot
signal ensures a stable, transient-free operation of the erbium-doped fibre amplifiers (EDFA).
Afterwards, the pilot signal is removed again by tunable bandpass filters. Then, the pulses
propagate through 4km of dispersion compensating fibre (DCF).
An additional single mode fibre (SMF) spool is needed to induce a length difference
of only $\delta L=50$m. During each roundtrip, 10\% of the intensity is removed
directly after the fibre spools by monitor couplers for pulse detection. For controlling the
polarization state within each loop, a polarising beam splitter (PBS) is inserted into the
left loop, while the phase modulators (PM) in each loop have an integrated polarizer.
Mechanical polarization controllers, marked by three black circles, allow for an adjustment
of the polarization at various places within the fibre loop. Besides the PMs in each loop,
which allow for an arbitrary phase shift, the acousto-optical modulators (AOM) are used to
modulate the amplitudes of the pulses in both loops.

\end{document}